\documentclass[aps,pra,prabib,twocolumn,showpacs,superscriptaddress]{revtex4} 
\usepackage{graphicx}
\usepackage{amsmath}
\usepackage{amssymb}
\usepackage{amsfonts}
\usepackage{bm}

\begin{document}

\title{Formation of a vortex lattice in a rotating BCS Fermi gas}

\affiliation{Dipartimento di fisica, Universit\`a di Firenze, Firenze, Italy}
\affiliation{Laboratoire Kastler Brossel, \'Ecole Normale
Sup\'erieure, 24 rue Lhomond, 75231 Paris Cedex 05, France}

\author{Giulia Tonini}
\affiliation{Dipartimento di fisica, Universit\`a di Firenze, Firenze, Italy}

\author{F\'elix Werner}
\affiliation{Laboratoire Kastler Brossel, \'Ecole Normale
Sup\'erieure, 24 rue Lhomond, 75231 Paris Cedex 05, France}
\author{Yvan Castin}
\email{yvan.castin@lkb.ens.fr}
\affiliation{Laboratoire Kastler Brossel, \'Ecole Normale
Sup\'erieure, 24 rue Lhomond, 75231 Paris Cedex 05, France}

\begin{abstract}
We investigate theoretically the formation of a vortex lattice in
a superfluid two-spin component Fermi gas in a rotating harmonic trap, in a BCS-type
regime of condensed non-bosonic pairs.
Our analytical solution of the superfluid
hydrodynamic equations, both for the 2D BCS equation of state
and for the 3D unitary quantum gas, predicts that the vortex free gas is 
subject to a dynamic instability for fast enough rotation. 
With a numerical solution of the full
time dependent BCS equations in a 2D model, we confirm the existence of this dynamic
instability and we show that it leads to the formation of a regular pattern
of quantum vortices in the gas.
\end{abstract}



\date{\today}

\maketitle

The field of trapped ultracold fermionic atomic gases is presently making rapid progress:
thanks to the possibility of controlling at will the strength of the $s$-wave interaction
between two different spin components by the technique of the Feshbach resonance
\cite{Thomas,Salomon_inter}, it is possible to investigate the cross-over \cite{CrossoverTheory}
between the weakly interacting
BCS regime (case of a small and negative scattering length) and the Bose-Einstein
condensation of dimers (case of small and positive scattering length), including the
strongly interacting regime and even the unitary quantum gas (infinite scattering length).
The interaction energy of the gas was measured on both sides of the Feshbach resonance
\cite{Salomon_inter}; 
on the side of the resonance with a positive scattering length, Bose-Einstein
condensation of dimers was observed \cite{MolecBEC}; on the side of
the resonance with a negative scattering length, a condensation of pairs was
revealed in the strongly interacting regime
by a fast ramping of the magnetic field across the Feshbach resonance
\cite{AtomicBCS}.  Also, the presence of a gap in the excitation 
spectrum was observed \cite{Grimm_gap}, for an excitation consisting in
transferring atoms to an initially empty atomic internal state, as initially
suggested by \cite{Zoller}, revealing pairing.

Are there evidences of superfluidity in these fermionic gases~? It was initially
proposed \cite{Menotti} to reveal superfluidity by detecting an hydrodynamic
behavior in the expansion
of the gas after a switching-off of the trapping potential. Such an hydrodynamic 
behavior was indeed observed \cite{Thomas} but it was then realized that this
can occur not only in the superfluid phase, but also in the
normal phase in the so-called hydrodynamic regime, that is when the mean free
path of atoms is smaller than the size of the cloud, a condition easy to fulfill
close to a Feshbach resonance. The general experimental trend is
now to try to detect superfluidity via an hydrodynamic behavior that has no
counterpart in the normal phase \cite{Stringari2}. 
A natural candidate to reveal superfluidity
is therefore the detection of quantum vortex lattices in the rotating trapped Fermi
gases: the superfluid velocity field, defined as the gradient of the phase
of the order parameter, is irrotational everywhere, except on singularities
corresponding to the vortex lines, so that a superfluid may respond to rotation
by the formation of a vortex lattice \cite{Feder}; 
on the contrary, a rotating hydrodynamic normal gas
is expected to acquire the velocity field of solid-body rotation and should not
exhibit a regular vortex lattice in steady state.

Steady state properties of vortices in a rotating Fermi gas described by BCS theory
have already been the subject of several studies, for a single
vortex configuration  \cite{Bruun} and more recently for a vortex
lattice configuration \cite{Feder}. In this paper, we study the issue of the time
dependent formation of the lattice in a rotating Fermi gas, by solving the time
dependent BCS equations. A central point of the paper is to identify 
possible nucleation mechanisms of the vortices that could show up
in a real experiment.

This problem was addressed a few years ago for rotating Bose gases. The expected
nucleation mechanism was the Landau mechanism, corresponding to the apparition
of negative energy surface modes for the gas in the rotating frame, for a rotation frequency
above a minimal value; these negative energy modes can then be populated thermally,
leading to the entrance of one or several vortices from the outside part of the trapped
cloud \cite{Machida,Dalfovo}. The first experimental observation of a vortex
lattice in a rotating Bose-Einstein condensate revealed however a nucleation
frequency different from the one of the thermal Landau mechanism \cite{Dalibard_vort}
and was suggested later on to be due to a dynamic instability of hydrodynamic
nature triggered by the rotating harmonic trap \cite{Sinha},
which was then submitted to experimental tests \cite{Dalibard_instab,Foot}.
The discovered mechanism of dynamic instability was checked, by a numerical solution
of the purely conservative time dependent Gross-Pitaevskii equation, 
to lead to turbulence \cite{Feder_turb} 
and to the formation of a vortex lattice \cite{Sinatra}.

We now transpose this problem to the case of a two spin component Fermi
gas, initially at zero temperature and stirred by a rotating harmonic 
trapping
potential of slowly increasing rotation speed, as described
in section \ref{sec:model}. Does the hydrodynamic instability
phenomenon occur also in the fermionic case, and does it lead to the entrance
of vortices in the gas and to the subsequent formation of a vortex lattice ?
We first address this problem analytically, in section \ref{sec:hydro},
by solving exactly the time dependent two-dimensional
hydrodynamic equations and by performing a linear stability analysis: very similarly
to the bosonic case, we find that a dynamic instability can occur above some
minimal rotation speed. We also extend this conclusion to the 3D unitary quantum gas.
Then we test this prediction by a numerical solution of the time dependent BCS
equations on a two-dimensional lattice model, in section \ref{sec:simul}: 
this confirms that the dynamic instability can take place 
and leads to the entrance of vortices in the gas, which are then seen
to arrange in a regular pattern at long evolution times.

\section{Our model}
\label{sec:model}

We consider a gas of fermionic particles of mass $m$, with equally populated two spin states 
$\uparrow$ and
$\downarrow$, trapped in a harmonic potential and initially at zero temperature. 
The particles with opposite spin have a $s$-wave interaction with a negligible
range interaction potential, characterized by the scattering length $a_{3D}$,
whereas the particles in the same spin state do not interact.

We shall be concerned mainly by the limit of a 2D Fermi gas.
In this case, the trapping potential is very strong
along $z$ axis so that the quantum of oscillation along $z$, that is $\hbar\omega_z$,
where $\omega_z$ is the oscillation frequency along $z$,  is much larger than both
the mean oscillation energy in the $x-y$ plane and the interaction energy per particle, so that
the gas is dynamically frozen along $z$ in the ground state of the corresponding harmonic
oscillator. 
In this geometry, the two-body  interaction can be characterized by the 2D scattering length $a_{2D}$
which was calculated as a function of the 3D scattering length
in \cite{Shlyapnikov}. We recall that $a_{2D}$ is always strictly positive 
and the 2D 
two-body problem in free space exhibits a bound state, that is a dimer, of spatial radius
$a_{2D}$. 
For the 2D gas to have universal many-body interaction properties, characterized by
$a_{\rm 2D}$ only,
one requires that the spatial extension $(\hbar/m\omega_z)^{1/2}$
of the ground state of the harmonic oscillator along $z$ is smaller than $a_{2D}$ 
\cite{Olshanii}, so that e.g.\ the dimer binding energy is smaller than
$\hbar\omega_z$.
The weakly attractive Fermi gas limit corresponds in 2D to $\rho a_{2D}^2
\rightarrow +\infty$
and the condensation of preformed dimers to $\rho a_{2D}^2\rightarrow 0$ \cite{Randeria_2d},
where $\rho$ is the 2D density of the gas.

In the $x-y$ plane, the zero temperature 2D gas is initially harmonically trapped in the non-rotating, anisotropic potential
\begin{equation}
U(\mathbf{r}) = \frac{1}{2} m \omega^2\left[(1-\epsilon) x^2 + (1+\epsilon) y^2\right]
\end{equation}
where $\mathbf{r}=(x,y)$ and $\epsilon>0$ measures the anisotropy of the trapping 
potential. Then one gradually sets the trapping potential into rotation around $z$ axis
with an instantaneous rotation frequency $\Omega(t)$, until it reaches a maximal value
$\Omega$ to which it then remains equal. The question is to study the resulting evolution of the
gas and predict the possible formation and subsequent crystallization of quantum vortices.

We shall consider this question within the approximate frame of the BCS theory, in a rather 
strongly interacting regime but closer to the weakly interacting 
BCS limit than to the BEC limit, which is most relevant for the present 3D experimental investigations: 
the chemical potential $\mu$ of the 2D gas is supposed
to be positive, excluding the regime of Bose-Einstein condensation of the dimers, 
and  the parameter $k_F a_{2D}$, where the Fermi momentum is
defined as $\hbar^2 k_F ^2/2m = \mu$, is larger than unity but not extremely larger than unity:
we shall take $k_F a_{2D}=4$ in the numerical simulations.
In this relatively strongly interacting regime, we of course do not expect the BCS theory to be 100\%
quantitative.

In the hydrodynamic approach to come, one simply needs the equation of state of the gas,
that is the expression of the chemical potential $\mu_0$ of a spatially uniform zero
temperature gas as a function of the total density $\rho=\rho_\uparrow+\rho_\downarrow=
2\rho_\uparrow$ and of the scattering length. In 2D, this equation of state was calculated
with the BCS approach in \cite{Randeria_2d}:
\begin{equation}
\mu_0[\rho]= \frac{\pi \hbar ^2\rho}{m} - E_0/2
\label{eq:eos}
\end{equation}
where $E_0$ is the binding energy of the dimer in free space, 
\begin{equation}
E_0=\frac{4\hbar^2}{m a_{2D}^2 e^{2\gamma}}
\end{equation}
and $\gamma= 0.57721\ldots$ is Euler's constant. Similarly, the gap for the zero temperature
homogeneous BCS gas is related to the density by \cite{Randeria_2d}
\begin{equation}
\Delta_0[\rho]= \left(E_0 \frac{2\pi \hbar ^2\rho}{m}\right)^{1/2}.
\label{eq:delta_randeria}
\end{equation}
We shall also consider analytically the 3D unitary quantum gas
($a_{3D}=\infty$) where the equation of state is known to be exactly of
the form $\mu_0[\rho]\propto \hbar^2\rho^{2/3}/m$.

In the numerical solution of the 2D time dependent BCS equations to come, one needs an explicit 
microscopic model. We have chosen a square lattice model with an on-site interaction between
opposite spin particles corresponding to a coupling constant $g_0$ so that the second
quantized grand canonical Hamiltonian reads at the initial time
\begin{eqnarray}
H &=& \sum_{\mathbf{k}, \sigma}  \left(\frac{\hbar ^2 k^2}{2m}-\mu\right) c_{\mathbf{k}, \sigma}^\dagger
c_{\mathbf{k}, \sigma} + \sum_{\mathbf{r},\sigma}  l^2\, U(\mathbf{r}) 
\psi_\sigma^\dagger(\mathbf{r}) \psi_\sigma(\mathbf{r}) \nonumber \\
&&+g_0 \sum_{\mathbf{r}} l^2 \, \psi_\uparrow^\dagger(\mathbf{r}) \psi_\downarrow^\dagger(\mathbf{r})
\psi_\downarrow(\mathbf{r}) \psi_\uparrow(\mathbf{r})
\end{eqnarray}
where $l$ is the grid spacing.
In the numerics a quantization volume is introduced, in the form of
a square box of size $L$ with periodic boundary conditions, $L$ being an integer multiple of $l$.
The sum over $\mathbf{r}$ then runs over the $(L/l)^2$ points of the lattice. A plane wave on
the lattice has wavevector components $k_x$ and $k_y$ having a meaning modulo $2\pi/l$ so that
the wavevector $\mathbf{k}$ is restricted to the first Brillouin zone $D=[-\pi/l,\pi/l[^2$.
The operator $c_{\mathbf{k}, \sigma}$ annihilates a particle of wavevector $\mathbf{k}$
and spin state $\sigma=\uparrow$ or $\downarrow$, and obeys the usual fermionic anticommutation
relations, such as
\begin{equation}
\{ c_{\mathbf{k}, \sigma}, c_{\mathbf{k'}, \sigma'}^\dagger\} = \delta_{\mathbf{k}, \mathbf{k'}}
\delta_{\sigma,\sigma'}.
\end{equation}
The discrete field operator $\psi_{\sigma}(\mathbf{r})$ is proportional to the annihilation operator
of a particle at the lattice node $\mathbf{r}$ in the spin state $\sigma$ in such a way that it obeys
the anticommutation relations
\begin{equation}
\{\psi_\sigma(\mathbf{r}), \psi_{\sigma'}^\dagger(\mathbf{r'})\} = 
l^{-2} \, \delta_{\mathbf{r},\mathbf{r'}} \delta_{\sigma,\sigma'}.
\end{equation}
The coupling constant $g_0$ is adjusted so that the 2D scattering length of two particles on 
the infinite lattice is exactly $a_{2D}$ \cite{Mora, Houches2003}:
\begin{equation}
\frac{1}{g_0} =  \frac{m}{2\pi\hbar^2} 
\left[\log \left(\frac{l}{\pi a_{2D}}\right)-\gamma + \frac{2 G}{\pi}
\right]
\end{equation}
where $G=0.91596\ldots $ is Catalan's constant. 
In the limit $a_{2D}\rightarrow +\infty$, for a fixed density $\rho$
and a fixed `range' $l$ of the interaction
potential, one finds $g_0\rightarrow 0^-$:  we recover the fact that the limit
$k_F a_{2D}\gg 1$ corresponds to a weakly attractive Fermi gas.

At later times, the Hamiltonian is written in the frame rotating at frequency $\Omega(t)$, to eliminate
the time dependence of the trapping potential; this adds an extra term to the Hamiltonian, 
\begin{equation}
H_{\rm rot} = -\Omega(t) \sum_{\mathbf{r},\sigma}  l^2 \psi_\sigma^\dagger(\mathbf{r}) 
\left(L_z \psi_\sigma\right)(\mathbf{r})
\end{equation}
where the matrix $L_z$ on the lattice represents the angular momentum operator along $z$,
$x p_y-y p_x$. The square box defining the periodic boundary conditions is supposed to be fixed
in the rotating frame, so that it rotates in the lab frame: this may be useful in practice to ensure 
that truncation effects due to the finite size of this box in the numerics do not arrest the rotation of the gas.

This lattice model is expected to reproduce a continuous model with harmonic trapping and zero range interaction potential
in the limit of an infinite quantization volume ($L\gg$ spatial radius of the cloud) and in the limit
of a vanishing grid size $l\rightarrow 0$ ($l\ll a_{2D}, k_F^{-1}$). 
In this limit $g_0$ is negative, 
leading to an attractive interaction, so that pairing can take place in the lattice model.
In this limit, we have checked that BCS theory for the lattice model gives the same equation
of state as Eq.(\ref{eq:eos}) \cite{these_Giulia}.

\section{Solution to the superfluid hydrodynamic equations}
\label{sec:hydro}

In the hydrodynamic theory of a pure superfluid with no vortex, one introduces two fields, the 
total spatial density
of the gas,  $\rho(\mathbf{r},t)$, and the phase
of the so-called order parameter, $2\, S(\mathbf{r},t)/\hbar$. In the BCS theory for the lattice model, the order
parameter is simply
\begin{equation}
\Delta(\mathbf{r},t) \equiv -g_0 \langle \psi_\uparrow(\mathbf{r},t) \psi_\downarrow(\mathbf{r},t) \rangle
\equiv |\Delta| e^{2 i S/\hbar}
\label{eq:delta}
\end{equation}
which has a finite limit when $l\rightarrow 0$.
The superfluid velocity field in the lab frame is then defined as
\begin{equation}
\mathbf{v} = \frac{\mathbf{grad}\, S}{m}.
\end{equation}

In the rotating frame, the hydrodynamic equations read
\begin{eqnarray}
\label{eq:continuity}
\partial_t \rho &=& - \mathrm{div}\, \left[\rho \left(\mathbf{v}-\mathbf{\Omega}(t)\times
\mathbf{r}\right)\right] \\
-\partial_t S &=& \frac{1}{2} m v^2 + U(\mathbf{r}) + \mu_0[\rho(\mathbf{r},t)]  \nonumber\\
&& -\mu - m(\mathbf{\Omega}(t) \times \mathbf{r})\cdot \mathbf{v}
\label{eq:euler}
\end{eqnarray}
where $\mathbf{\Omega}(t)=\Omega(t)\hat{\mathbf{z}}$ and $\hat{\mathbf{z}}$ is the unit vector along the rotation axis $z$. The first equation is simply the continuity
equation in the rotating frame, including the fact
that the velocity field in the rotating frame differs from the one
in the lab frame by the solid body  rotational term. When one takes the gradient of the second equation,
one recovers Euler's equation for a superfluid. These superfluid equations are expected to be correct
for a slowly varying density and phase, both in space (as compared to the size of a BCS pair)
and in time (as compared to $\hbar/|\Delta|$). 
For a harmonically trapped system with a quantum of oscillation $\hbar\omega$, 
the slow spatial variation condition requires
a gap parameter $|\Delta|\gg \hbar\omega$:
in the present paper, considering the rather strongly
interacting regime $1\lesssim k_F a_{2D}$, the gap is of the order of the Fermi energy,
which is much larger than $\hbar\omega$, 
so that there is  slow spatial variation as long
as no vortex enters the cloud. The gap is then much larger than $\hbar$ over
the ramping time of the trap rotation, 
so that the expected condition of slow time variation is also satisfied.
In the appendix \ref{appen:hydro} we present a simple but systematic
derivation of these superfluid hydrodynamic equations 
starting from the time dependent BCS theory and using a 
semi-classical expansion.
Surprisingly, for the case of slow ramping times and rather fast rotations considered in this paper,
with $\Omega$ of the order of $\omega$,
our simple derivation requires an extra validity condition, in general more stringent than
$|\Delta|\gg\hbar\omega$: the quantum of oscillation
$\hbar \omega$ should be smaller than $|\Delta|^2/\mu$, a condition also satisfied in our simulations.

We shall assume here that the rotation frequency is ramped up very slowly so that the density and
the phase adiabatically follow a sequence of vortex free stationary states. The strategy then
closely follows the one already developed in the bosonic case \cite{Sinha}: one solves analytically
the corresponding stationary hydrodynamic equations, then one performs a linear stability analysis
of the stationary solution. The apparition of a dynamic instability suggests that the system may
evolve far away from the stationary branch; that this dynamic instability results in the entrance
of vortices will be checked by the numerical simulations of section \ref{sec:simul}.

\begin{figure}[htb]
\begin{center}
\includegraphics[width=8cm,clip]{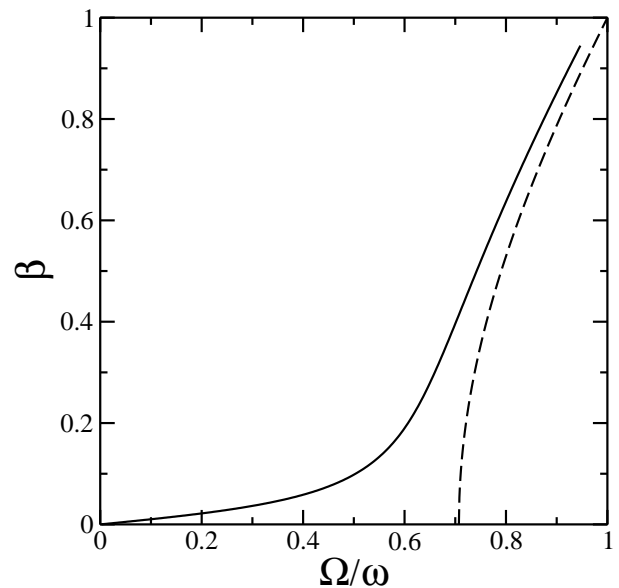}
\end{center}
\caption{The upper branch of solution for the phase parameter $\beta$ of the hydrodynamic approach
for a stationary vortex free BCS state in the rotating frame, as a
function of the rotation frequency. Solid line: the trap anisotropy is $\epsilon=0.1$.
Dashed line: $\epsilon=0$.}
\label{fig:beta}
\end{figure}

In the stationary regime, for a fixed rotation frequency $\Omega$,
one sets $\partial_t \rho=0$ in Eq.(\ref{eq:continuity})
and $-\partial_t S=0$ in Eq.(\ref{eq:euler}) \cite{proviso}. 
We first consider the 2D case and we replace $\mu_0$ by the equation of state
Eq.(\ref{eq:eos}): apart from an additive constant, $\mu_0$ is proportional to the density,
as was the case for the weakly interacting condensate of bosons \cite{Sinha}, so that the calculations
for the superfluid fermions are formally the same, if one replaces the coupling constant
$g$ of the bosons by $\pi\hbar^2/m$. Since the properties of the bosons do not depend on the value of $g$
up to a scaling on the density \cite{Sinha}, the results for the bosons can be directly transposed. 
Following \cite{Stringari_alpha}, we take the ansatz
for the phase:
\begin{equation}
S(\mathbf{r}) =  m \omega \beta x y
\label{eq:ansatz_xy}
\end{equation}
which is applicable for a harmonic trapping potential $U$.
When inserted in Eq.(\ref{eq:euler}), this leads to an inverted parabola for the density profile, resulting
in an elliptic boundary for the density of the cloud.
Upon insertion of the density profile in the continuity equation, one recovers the cubic
equation of \cite{Stringari_alpha}:
\begin{equation}
\label{eq:cubic}
\beta^3 +\left(1-2\frac{\Omega^2}{\omega^2}\right)\, \beta - \epsilon \frac{\Omega}{\omega} = 0.
\end{equation}
This equation has one real root for $\Omega$ below some $\epsilon$ dependent
bifurcation value $\Omega_{\rm bif}(\epsilon)$, and has three real roots for $\Omega>
\Omega_{\rm bif}(\epsilon)$. 
In the considered stirring procedure, the system
starts with $\beta=0$ and follows adiabatically the so-called 
upper branch of solution, corresponding
to increasing values of $\beta$. 
In figure \ref{fig:beta}, we have plotted $\beta$ 
as a function of $\Omega/\omega$ on this branch, 
for the value of the asymmetry parameter in the simulations
of the next section,
$\epsilon=0.1$. 
When $\beta$ takes appreciable values, the cloud significantly deforms itself
in real space, becoming broader along $x$ axis than along $y$ axis, 
even for an arbitrarily weak trap anisotropy
$\epsilon$.

From the studies of the bosonic case \cite{Sinha} it is known that the significantly deformed clouds 
can become dynamically unstable. We recall briefly the calculation procedure: one introduces initially
arbitrarily small deviations $\delta \rho$ and $\delta S$ of the density and the phase from
their stationary values; one then linearizes the hydrodynamic equations Eq.(\ref{eq:continuity})
and Eq.(\ref{eq:euler}) to get
\begin{eqnarray}
\label{eq:delta_S}
\frac{D\,\delta\rho}{Dt} &=& - \mathrm{div} \, \left(\rho\frac{\mathbf{grad}\,\delta S}{m}\right)
\\
\frac{D\,\delta S}{Dt} &=& -\frac{\pi\hbar^2}{m} \, \delta \rho
\end{eqnarray}
where $D/Dt\equiv \partial_t +
(\mathbf{v}-\mathbf{\Omega}\times\mathbf{r})\cdot\mathbf{grad}$ and where
we used the fact that the Laplacian of $S(\mathbf{r})\propto xy$ vanishes. 
One then calculates the eigenmodes of the linearized equations, setting
$\partial_t \rightarrow -i \nu$ where $\nu$ is the eigenfrequency of the mode.
As an ansatz for $\delta\rho(\mathbf{r})$ and $\delta S(\mathbf{r})$, one
takes polynomials of arbitrary total degree $n$ in the variables $x$ and
$y$. One can indeed check that the subspace of polynomials
of degree $\leq n$ is stable, since the stationary values $\rho$ and $S$ are quadratic
functions of $x$ and $y$.
This turns the linearized partial differential equations into a finite size
linear system whose eigenvalues can be calculated numerically.
Complex eigenfrequencies, when obtained, lead to a non-zero Lyapunov exponent
$\lambda\equiv \mathrm{Im}\,\nu$, which reveals a dynamical instability when $\lambda >0$.

\begin{figure}[htb]
\begin{center}
\includegraphics[width=8cm,clip]{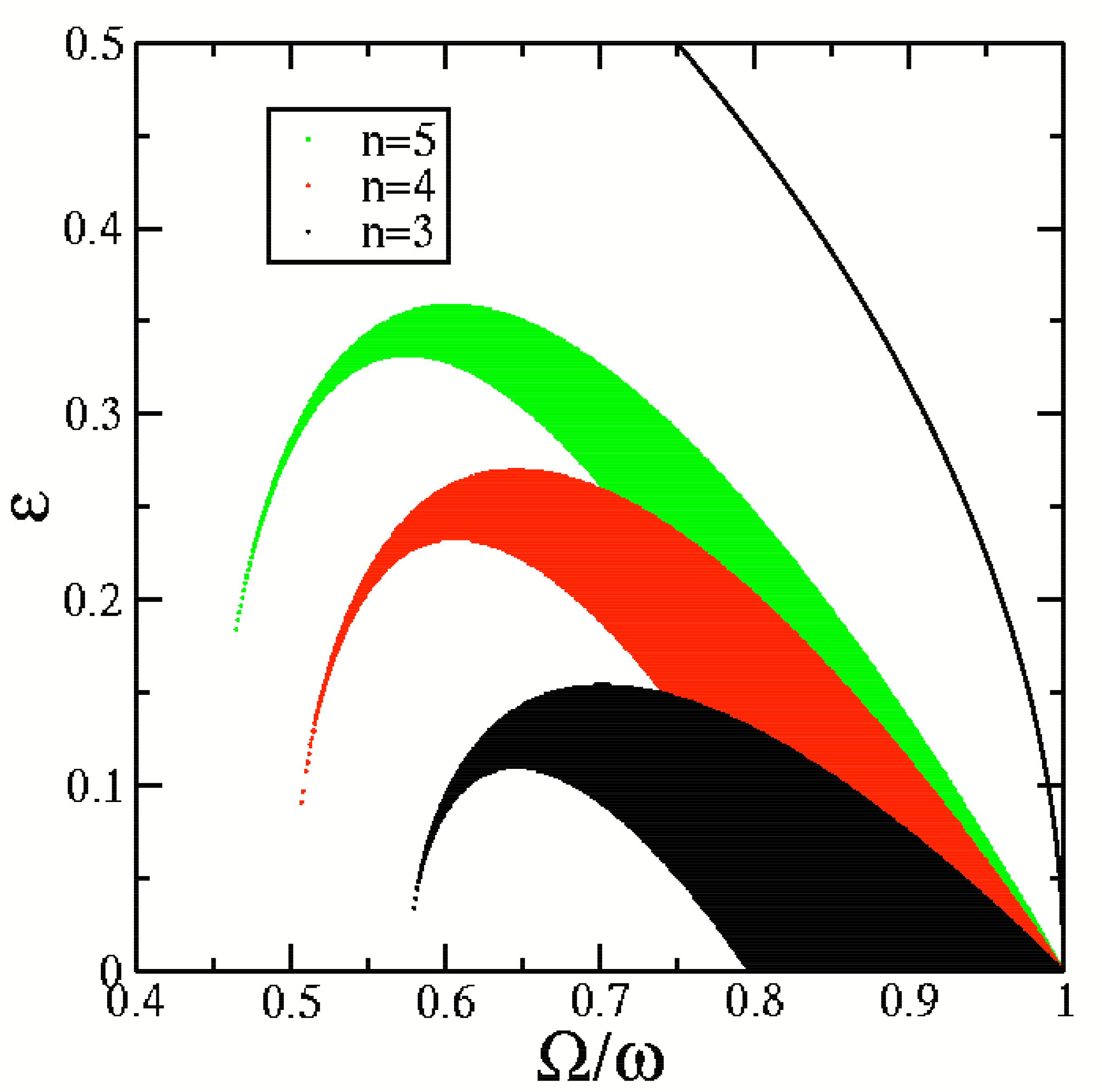}
\end{center}
\caption{For the upper branch of solution for the phase parameter, in 2D:
Dark areas: instability domain in the $\Omega-\epsilon$ plane 
for degrees $n$ equal to 3, 4 and 5 (crescents from bottom to top).
There is no dynamical instability for $n\leq 2$. Solid line: border
$\Omega^2= (1-\epsilon)\omega$ of the branch existence domain.}
\label{fig:diagram}
\end{figure}

\begin{figure}[htb]
\begin{center}
\includegraphics[width=8cm,clip]{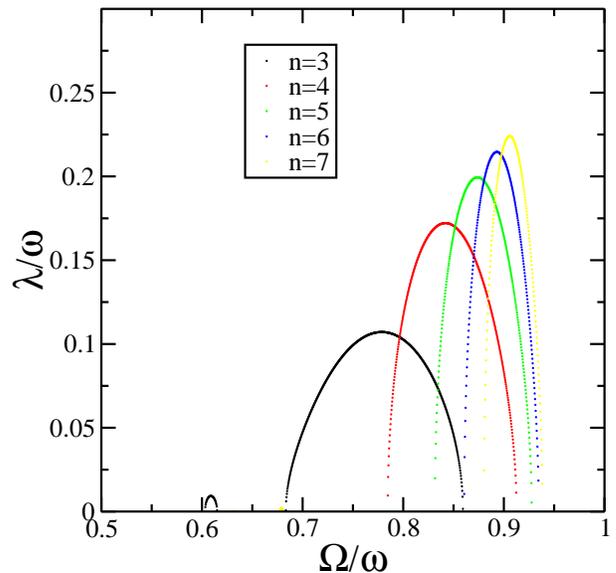}
\end{center}
\caption{For the upper branch of solution for the phase parameter in 2D:
Lyapunov exponent of the dynamic instability for degrees $n$ from 3 to 7, as
a function of the rotation frequency. The trap anisotropy is
$\epsilon=0.1$.}
\label{fig:lambda}
\end{figure}

In figure \ref{fig:diagram} we plot the stability diagram of the upper branch stationary solution
in the plane $(\Omega,\epsilon)$, for various total degrees $n$ of the polynomial ansatz. Each degree contributes
to this diagram in the form of a crescent, touching the horizontal axis ($\epsilon=0$)
with a broad basis on the right side and a very narrow tongue on the left side
\cite{to_be_complete}.
For the low value $\epsilon=0.1$ considered in the numerical simulations of this paper,
the Lyapunov exponents in the tongues are rather
small, so that significant instability exponents are found only in the broad bases: 
for increasing $\Omega$, the first encountered significant instability corresponds to a degree $n=3$: for $\epsilon=0$, the corresponding minimal value
of $\Omega/\omega$ is $(\frac{183+36\sqrt{30}}{599})^{1/2}=0.79667\ldots$
\cite{how_to_get_that}.
This is apparent in figure \ref{fig:lambda}, where we plot the Lyapunov exponent as a function of $\Omega/\omega$
for various degrees $n$ and for $\epsilon=0.1$.

\noindent {\it Extension to the unitary quantum gas in 3D:}
In practice, the experiments are mainly performed in 3D, so that
we generalize the previous hydrodynamic calculation to a 3D case
where the exact equation of state is known: the so-called unitary regime,
where the 3D $s$-wave scattering length between opposite spin
fermions is infinite.
Because of the universality of the unitary quantum gas,
the equation of state of the gas is indeed a power law
\begin{equation}
\mu_0[\rho] = A \rho^\gamma
\end{equation}
where the exponent $\gamma=2/3$ and where the factor $A$ 
is proportional to $\hbar^2/m$, with a proportionality constant
recently calculated with fixed node Monte Carlo methods \cite{Pandha,Giorgini}
and measured in recent experiments by Grimm \cite{Grimm_eta}
and by Salomon \cite{MolecBEC}.

For such a non-linear equation of state, one seems to have lost
the underlying structure
of the hydrodynamic equations allowing a quadratic ansatz for $\rho$ and $S$, 
and a polynomial ansatz for $\delta\rho$ and $\delta S$. Fortunately,
this structure can be restored by using as a new variable
$R(\mathbf{r},t)\equiv \rho^\gamma(\mathbf{r},t)$.
One then gets effective hydrodynamic equations with a linear
equation of state:
\begin{eqnarray}
\label{eq:continuity_eff}
\partial_t R &=& - \gamma R\,\mathrm{div}\, 
\mathbf{v}
-(\mathbf{v}-\mathbf{\Omega}(t)\times\mathbf{r})\cdot
\mathbf{grad}\, R \\
-\partial_t S &=& \frac{1}{2} m v^2 + U_{3D}(\mathbf{r}) + A\, R(\mathbf{r})
 \nonumber\\
&& -\mu - m(\mathbf{\Omega}(t) \times \mathbf{r})\cdot \mathbf{v},
\label{eq:euler_eff}
\end{eqnarray}
where  the 3D trapping potential is 
\begin{equation}
U_{3D}(\mathbf{r}) = \frac{1}{2} m \omega^2\left[(1-\epsilon) x^2 + (1+\epsilon) y^2\right]+\frac{1}{2} m\omega_z^2 z^2.
\end{equation}
One then recycles the previous approach, with the usual quadratic ansatz for
the steady state values of $R$ and $S$. In particular
the same cubic equation for $\beta$ as in Eq.(\ref{eq:cubic}) is obtained. 
Linearizing the effective
hydrodynamic equations around the steady state, one gets
\begin{eqnarray}
\frac{D\,\delta R}{Dt} &=& - \gamma R 
\frac{\Delta_{\mathbf{r}} \delta S}{m}
-\frac{1}{m}\,\mathbf{grad}\,\delta S\cdot\mathbf{grad}\,R 
\label{eq:delta_S_eff} \\
\frac{D\,\delta S}{Dt} &=& - A \, \delta R,
\end{eqnarray}
where we used the fact that $S$ has a vanishing Laplacian.
This system of partial different equations
can be solved by a polynomial ansatz for $\delta S$ and
$\delta R$. This generalizes to the rotating case
the ansatz of \cite{Minguzzi}.

\begin{figure}[htb]
\begin{center}
\includegraphics[width=8cm,clip]{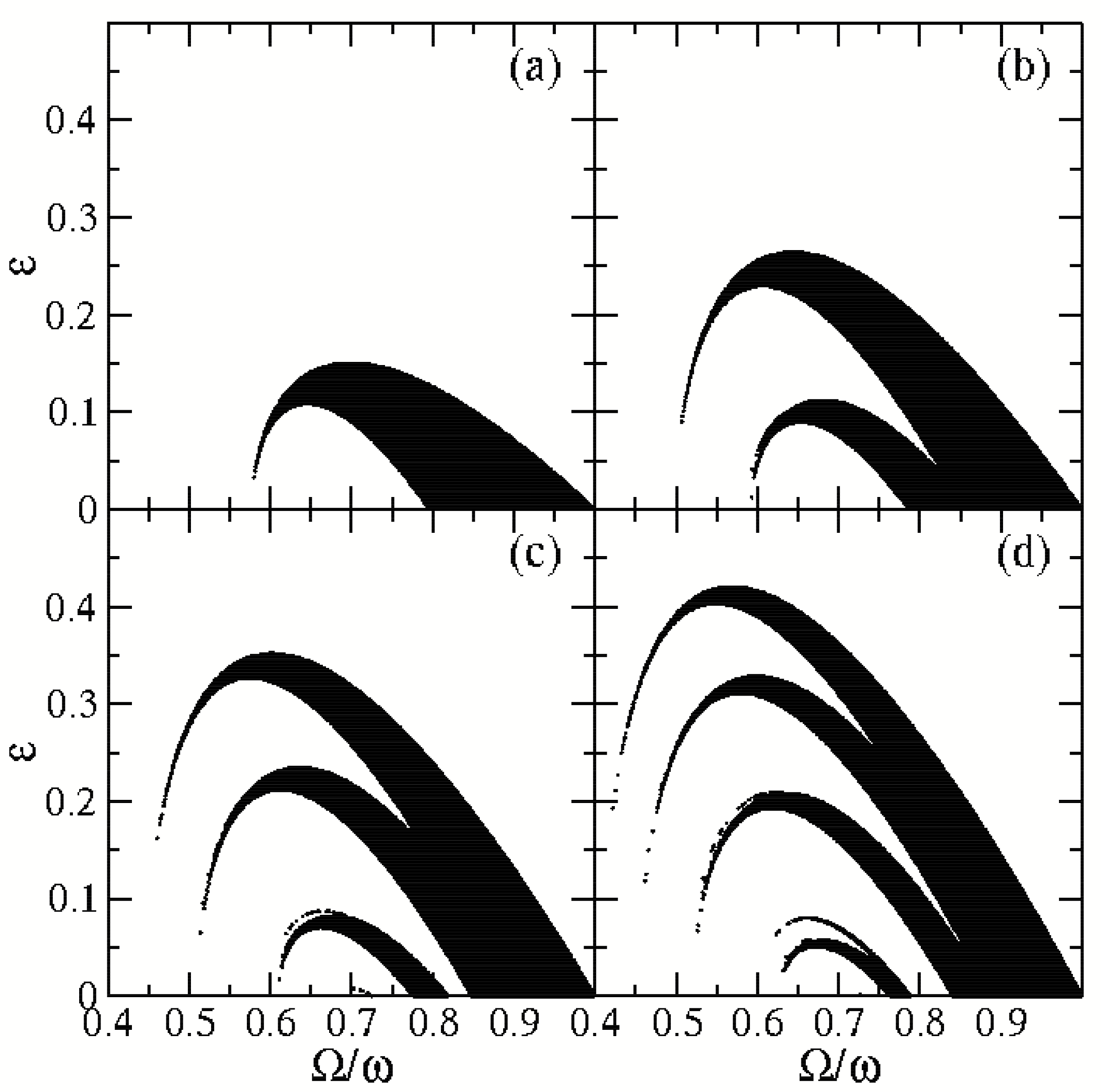}
\end{center}
\caption{Case of the 3D unitary quantum gas with
$\omega_z=0.4\omega$, for the upper branch of solution for the phase parameter:
Dark areas: instability domain in the $\Omega-\epsilon$ plane 
for degrees (a) $n=3$, (b) $n=4$, (c) $n=5$ and
(d) $n=6$.
There is no dynamical instability for $n\leq 2$.  }
\label{fig:3d}
\end{figure}

\begin{figure}[htb]
\begin{center}
\includegraphics[width=8cm,clip]{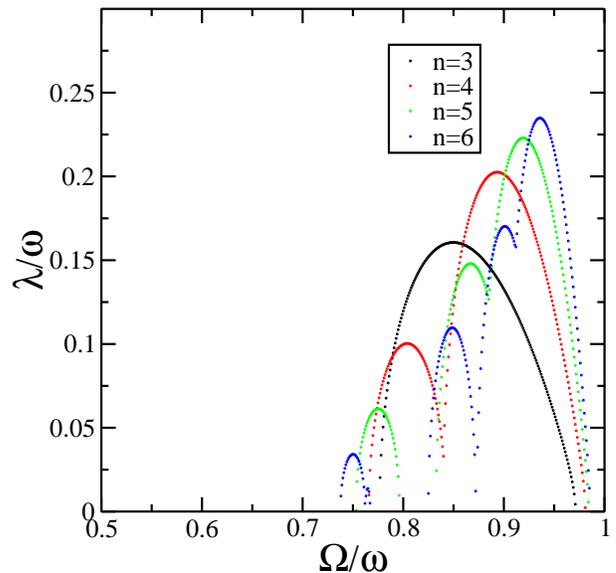}
\end{center}
\caption{Case of the 3D unitary quantum gas with $\omega_z=0.4\omega$,
for the upper branch of solution for the phase parameter:
Maximal
Lyapunov exponent of the dynamic instability for degrees $n$ from 3 to 6, as
a function of the rotation frequency. The trap anisotropy is
$\epsilon=0.022$.}
\label{fig:3dbis}
\end{figure}

In figure \ref{fig:3d} we have plotted the stability diagram of the upper 
branch stationary solution in the plane $(\Omega,\epsilon)$ for the
3D unitary quantum gas, for a trapping potential with
$\omega_z=0.4\omega$. The 3D nature of the problem makes
the structure of the instability domain more involved that in 2D.
This also appears in figure \ref{fig:3dbis}, giving the Lyapunov
exponents as a function of $\Omega$ for a fixed trap anisotropy
in the $x-y$ plane, $\epsilon=0.022$.
In the limit of a cigar shaped potential, $\omega_z\ll\omega$,
the structure is on the contrary close to the 2D one, as some of
the eigenmodes for $\delta R$ and $\delta S$ 
almost factorize in a function of $x,y$ and a function of $z$.

\section{Numerical solution of the 2D time dependent BCS equations}
\label{sec:simul}

We recall briefly the BCS equations for our two-component lattice model, in the case of equal
populations of the two spin states.
In the non-rotating case, the many-body ground state of the Hamiltonian
is approximated variationally in the zero temperature BCS theory 
by a so-called quasiparticle vacuum \cite{Blaizot}, 
that is the vacuum state of annihilation operators
of elementary excitations, $b_{s,\sigma}$ (where $\sigma=\uparrow$ or $\downarrow$).
By energy minimization, one finds that the $b_{s,\sigma}$ are such that
\begin{eqnarray}
\psi_\uparrow(\mathbf{r}) &=& \sum_s \left[b_{s,\uparrow} u_s(\mathbf{r}) -  b_{s,\downarrow}^\dagger v_s^*(\mathbf{r})\right] \label{eq:modal1} \\
\psi_\downarrow(\mathbf{r}) &=& \sum_s \left[b_{s,\downarrow} u_s(\mathbf{r}) +  b_{s,\uparrow}^\dagger v_s^*(\mathbf{r})\right] 
\label{eq:modal2}
\end{eqnarray}
where the $u$'s and $v$'s are all the eigenvectors of the following Hermitian system with positive energies $E_s>0$:
\begin{equation}
E_s\, \left(\begin{tabular}{c} $u_s$ \\ $v_s$\end{tabular}\right) = 
\left(\begin{tabular}{cc} $h_0$ & $\Delta$  \\ $\Delta^*$ & $-h_0^*$ \end{tabular}\right) 
\left(\begin{tabular}{c} $u_s$ \\ $v_s$\end{tabular}\right)
\label{eq:eigen}
\end{equation}
and normalized so that 
\begin{equation}
l^2 \sum_\mathbf{r} \left[|u_s(\mathbf{r})|^2+|v_s(\mathbf{r})|^2\right] = 1.
\end{equation}
In the eigensystem, $\Delta$ is the position dependent gap parameter defined in Eq.(\ref{eq:delta})
and the matrix $h_0$
represents on the lattice the single particle kinetic energy plus chemical potential plus harmonic potential energy terms.
When the modal decompositions Eqs.(\ref{eq:modal1},\ref{eq:modal2}) 
are inserted in Eq.(\ref{eq:delta}), one gets
\begin{equation}
\label{eq:gape}
\Delta(\mathbf{r}) = - g_0 \sum_s u_s(\mathbf{r}) v_s^*(\mathbf{r}).
\end{equation}
The density profile of the gas is given by
\begin{equation}
\label{eq:dens}
\rho(\mathbf{r}) = 2\langle \psi_\uparrow^\dagger(\mathbf{r}) \psi_\uparrow(\mathbf{r})\rangle
=2\sum_s |v_s(\mathbf{r})|^2.
\end{equation}
These equations actually belong to the zero temperature Hartree-Fock-Bogoliubov formalism
for fermions and are derived in \S 7.4b of \cite{Blaizot}.
Note that we have omitted the Hartree-Fock mean field term
\cite{mean_field}.

To solve numerically the 2D self-consistent stationary BCS equations, we have used the following
iterative algorithm: one starts with an initial guess for the position dependence of the
gap parameter (we used the local density approximation, taking advantage
of the fact that the equation of state Eq.(\ref{eq:eos}) 
and the value of the gap  Eq.(\ref{eq:delta_randeria}) within BCS
theory are known analytically
in 2D), then one calculates the $u$'s and $v$'s by diagonalization
of the Hermitian matrix in Eq.(\ref{eq:eigen}), one calculates the corresponding
$\Delta(\mathbf{r})$ 
using Eq.(\ref{eq:gape}), and one iterates until convergence.

Once the stationary BCS state is calculated, one moves to the solution of the 2D time
dependent BCS equations, to calculate the dynamics in the rotating trap.
What we call here time dependent BCS theory is the time-dependent Hartree-Fock-Bogoliubov
formalism for fermions, in the form of a variational calculation with a time
dependent quasiparticle vacuum $|\phi(t)\rangle$, as detailed in \S 9.5 of \cite{Blaizot}.
At time $t$, the modal expansions Eqs.(\ref{eq:modal1},\ref{eq:modal2}) still hold 
for $\psi_{\uparrow}(\mathbf{r})$ and $\psi_\downarrow(\mathbf{r})$, except that the
operators $b_{s,\sigma}$ (where $\sigma=\uparrow$ or $\downarrow$)  and the
mode functions are now time dependent. The variational state vector
$|\phi(t)\rangle$ is the vacuum of all the operators $b_{s,\sigma}(t)$.
The mode functions evolve according to
\begin{equation}
i\hbar \partial_t \, \left(\begin{tabular}{c} $u_s$ \\ $v_s$\end{tabular}\right) = 
\left(\begin{tabular}{cc} $h_0$ & $\Delta$  \\ $\Delta^*$ & $-h_0^*$ \end{tabular}\right) 
\left(\begin{tabular}{c} $u_s$ \\ $v_s$\end{tabular}\right)
\label{eq:uvt}
\end{equation}
where $h_0$ now includes the rotational term  $-\Omega(t) L_z$ in addition to the kinetic energy,
the chemical potential and the trapping potential. The gap function $\Delta$ is still
given by Eq.(\ref{eq:gape}) and is now time dependent as the mode functions are.
Note that Eq.(\ref{eq:uvt}) corresponds to the first of the equations (9.63b) in
\S 9.5 of \cite{Blaizot}, up to a global complex conjugation.
To be complete, we give the expression of the time dependent quasiparticle annihilation
operators:
\begin{eqnarray}
b_{s,\uparrow}(t) &=& l^2 \sum_{\mathbf{r}} u_s^*(\mathbf{r},t) \psi_\uparrow(\mathbf{r})
+v_s^*(\mathbf{r},t) \psi_\downarrow^\dagger(\mathbf{r}) \\
b_{s,\downarrow}(t) &=& l^2 \sum_{\mathbf{r}} u_s^*(\mathbf{r},t) \psi_\downarrow(\mathbf{r})
-v_s^*(\mathbf{r},t) \psi_\uparrow^\dagger(\mathbf{r}).
\end{eqnarray}
We also recall that this time-dependent formalism contains
not only pair-breaking excitations, but also implicitly
collective modes of the gas, as can be shown by a linearization
of these equations around a steady-state solution, see \S 10.2 in
\cite{Blaizot}, and as also shown by the fact that hydrodynamic equations
may be derived from them as done in the Appendix \ref{appen:hydro}. 
The numerical simulations to come therefore include excitations
of these collective modes, when the numerical solution deviates from a 
stationary state.

We have integrated numerically Eq.(\ref{eq:uvt}). The usual FFT split technique, which
exactly preserves the orthonormal nature of the $u$'s and $v$'s, is actually not satisfactory
because it assumes that the gap function remains constant in time during one time
step, which breaks the self-consistency of the equations and leads to a violation
of the conservation of the mean number of particles. We therefore used an improved
splitting method detailed in the appendix \ref{appen:split}.

In all the simulations that we present in this paper,
the trap anisotropy was $\epsilon=0.1$,
the chemical potential of the initial state of the gas
was fixed to $\mu=8\hbar \omega$; setting $\mu=\hbar^2 k_F ^2/2m$, the 2D scattering
length was fixed to the value $a_{\rm 2D}=(\hbar/m\omega)^{1/2}
\equiv a_{\rm ho}$
such that $k_F a_{\rm 2D} =4$; the rotation frequency was
turned on with the following law
\begin{equation}
\Omega(t)  = \Omega \, \sin^2\left(\frac{\pi t}{2\tau}\right) \ \ \ \ \mbox{for} \ \ 0\leq t\leq \tau
\end{equation}
with a ramping time $\tau=160 \omega^{-1}$ much larger than the oscillation period
of the atoms in the trap. For $t>\tau$, the rotation frequency remains equal to $\Omega$.
The presence of vortices is detected by calculating the winding number of the
phase of the gap parameter around each plaquette of the grid. We also calculated
the total angular momentum of the gas.
In all the simulations, we evolved the system for a total time of $1000 \, \omega^{-1}$. 

\noindent {\it Simulations on a small $32\times 32$ grid:}
For such a grid size, the calculation time remains reasonable so that
we varied the final rotation frequency
in steps of $0.1 \omega$. For final rotation frequencies
$\Omega \leq 0.3\omega$, no vortices are found to enter the cloud and the cloud remains round.

\begin{figure}[htb]
\begin{tabular}{cc}
\includegraphics[width=43mm,clip]{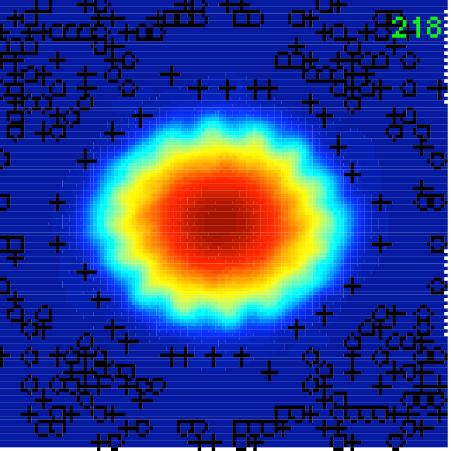} &
\includegraphics[width=43mm,clip]{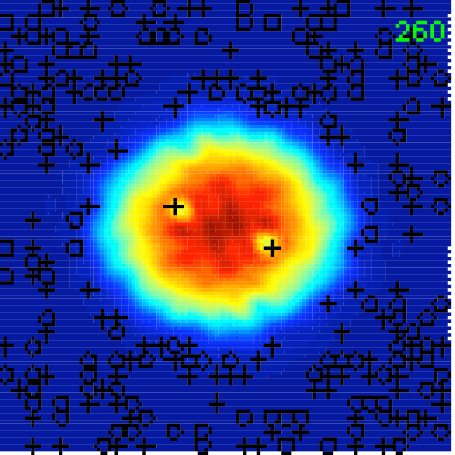} \\

\includegraphics[width=43mm,clip]{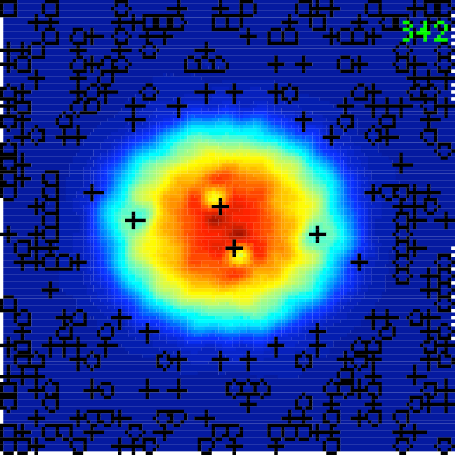} &
\includegraphics[width=43mm,clip]{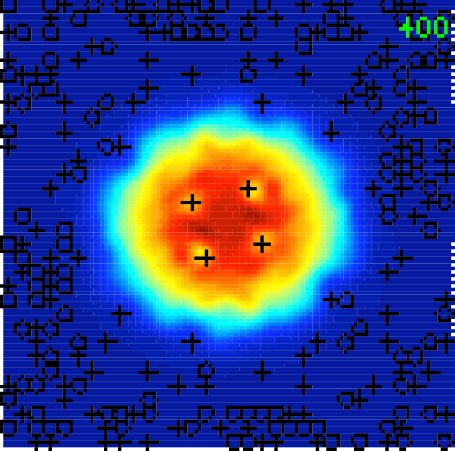} \\

\includegraphics[width=43mm,clip]{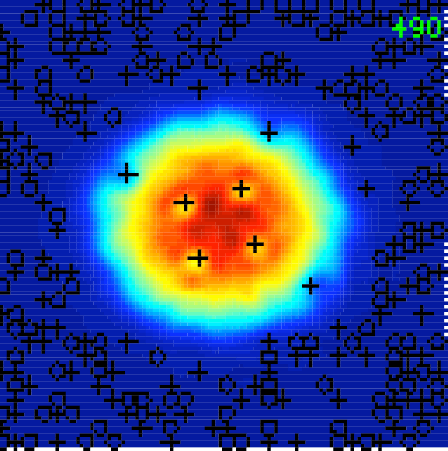} &
\includegraphics[width=43mm,clip]{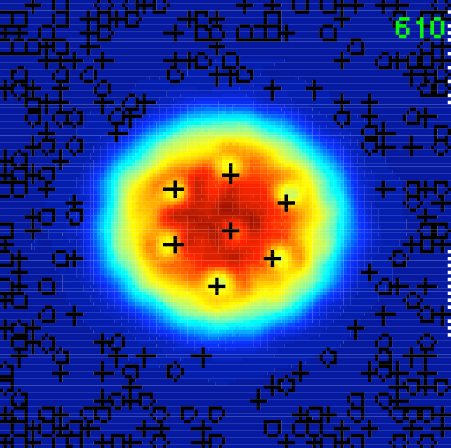} \\

\end{tabular}
\caption{For the numerical simulation of the 2D time dependent BCS equations on
a $32\times 32$ grid, density plots
of the density of the trapped gas at selected times (in units of $\omega^{-1}$), 
for a final rotation frequency $\Omega=0.5\omega$.
The trap anisotropy was $\epsilon=0.1$ and the 2D scattering length
$a_{2D}=\sqrt{\hbar/m\omega}$,
and $\mu=8\hbar\omega$ in the initial state.
The full spatial width of the simulation grid is shown in the figure.
Crosses: positive charge vortices.
Circles: negative charge vortices.
From top to bottom and from left to right:
$t=218\omega^{-1}$: a corrugation of the surface appears; $t=260\omega^{-1}$: a vortex pair enters;
$t=342\omega^{-1}$: a second vortex pair enters; $t=400\omega^{-1}$: the vortices arrange on a square;
$t=490\omega^{-1}$: a triplet of vortices starts entering; $t=610\omega^{-1}$: a stationary 7-vortex
pattern.
}
\label{fig:movie_0.5}
\end{figure}

For $\Omega=0.4\omega$, the cloud remains round but a corrugation of the surface of the cloud is
observed to appear at time $t\simeq 240 \omega^{-1}$; the amplitude of the corrugation
increases and two diametrically opposite vortices enter the cloud gently at $t\simeq 400
\omega^{-1}$ and, after a time interval of $\sim 100\omega^{-1}$,
settle in a stationary
pair of vortices close to the trap center. At $t\simeq 700 \omega^{-1}$, a second pair
of vortices starts entering with the same mechanism; it then interacts with the first
pair. The 4 vortices arrange in a stationary square at $t\simeq 850\omega^{-1}$.
For $\Omega=0.5\omega$, the situation is similar: one vortex pair enters, then a second one,
then a triplet of vortices starts entering at $t\simeq 490\omega^{-1}$; eventually, at
$t>610\omega^{-1}$ the 
seven vortices arrange in a stationary regular pattern, consisting of an hexagon
plus a vortex in the center.
Selected images of the movie of the simulation for
$\Omega=0.5\omega$ are shown in figure \ref{fig:movie_0.5}.
For $\Omega=0.6\omega$, the scenario is slightly different. The corrugation of the surface
is stronger, and a rectangular pattern of 4 vortices enter at time
$t\simeq 220\omega^{-1}$, shortly followed at $t\simeq 250 \omega^{-1}$ by
a second rectangular pattern of 4 vortices. After an interaction period, 6 vortices
align in the cloud in two parallel rows whereas two vortices are pushed away.
Then a third rectangle of vortices enter. At later times, several extra vortices
join the group; from $t\simeq 500 \omega^{-1}$ til the end of the simulation, 
12 vortices are present in the cloud, forming an almost stationary and regular pattern.
Clearly, in these scenarios, no global turbulence of the cloud is involved, since the first entering
vortices are arranged in a preformed pattern obeying the parity symmetry of the Hamiltonian.

For $\Omega=0.7\omega$ and $\Omega=0.8\omega$ the dynamics is very different from the
previous one.
The shape of the cloud strongly elongates and deforms.
Then strong turbulence sets in, at $t\simeq 160 \omega^{-1}$ for $\Omega=0.7\omega$ 
($t\simeq 135 \omega^{-1}$ for $\Omega=0.8\omega$), while the cloud anisotropy reduces,
the density profile becomes irregular, not only close the cloud boundary but also in the cloud center;
one observes a quick entrance of disordered vortices in the cloud at time $t\simeq 190 \omega^{-1}$
for $\Omega=0.7\omega$ ($t\simeq 150\omega^{-1}$ for $\Omega=0.8\omega$); 
several anti-vortices reach the borders of the cloud for $\Omega=0.7\omega$
and even reach high density regions for $\Omega=0.8\omega$.
After some evolution time, the density profile recovers a smooth and elliptic
shape, the anti-vortices are expelled from the cloud and the vortex positions slowly relax
to form a 17 (or 25 for $\Omega=0.8\omega$) vortex `lattice' at times $\sim 500\omega^{-1}$,
that remains essentially stationary til the end of the simulation. 

In conclusion, two distinct scenarios of vortex lattice formation are observed
in the $32\times 32$ simulations. For the lower rotation frequencies, a gentle entry
of an ordered pattern of vortices is observed. For the higher rotation frequencies, 
turbulence sets in and leads to the abrupt and disordered entrance of vortices
and even anti-vortices, the regular and stationary
vortex `lattice' forming after some  evolution time.
Another difference between the two scenarios is the temporal behavior of the density
profile at the location of the vortex cores: whereas a dip in the density is
visible from the start when a vortex enters the cloud with the gentle
scenario, such a dip at the vortex location 
forms only after some relaxation time in the turbulent scenario.

The physical origin of the turbulent scenario is expected to be the dynamic 
instability of the mode of degree $n=3$ discussed in section \ref{sec:hydro}, and the
obtained movies qualitatively agree with that. More quantitatively:
for $\epsilon=0.1$ a significant Lyapunov exponent is obtained for $\Omega>0.68\omega$;
this is compatible with the fact that the numerical simulation observes turbulence
for $\Omega\geq 0.7\omega$ only.

\begin{figure}[htb]
\begin{center}
\includegraphics[width=8cm,clip]{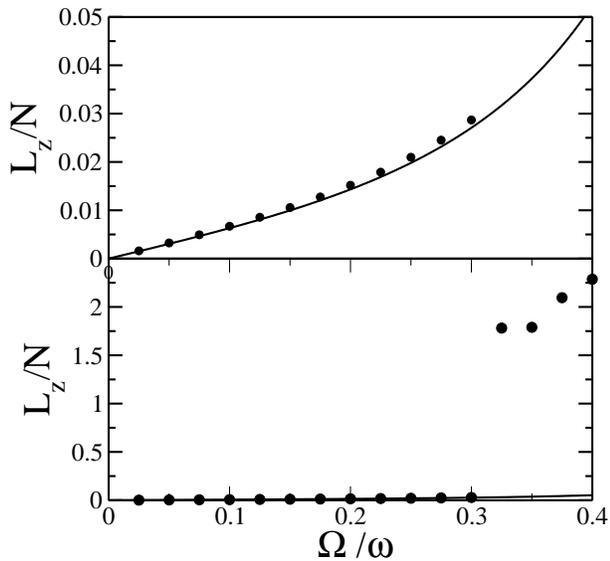}
\end{center}
\caption{In a steady state solution of the 2D BCS theory, on
a $32\times 32$ grid, angular momentum per particle in the gas, 
in units of $\hbar$,
as a function of the rotation frequency, for
$\epsilon=0.1$ and $\mu=8\hbar \omega$, $a_{2D}=(\hbar/m\omega)^{1/2}$. 
Black disks: numerical result from an iterative algorithm
(no vortex at the left part of the jump, 
vortices present at the right part of the jump).
Solid line: hydrodynamic prediction
(no vortex). The upper graph corresponds to the same data as
the lower graph, but for a different scale of the vertical axis:
it shows the good agreement of the hydrodynamic prediction
with the full numerics. In the numerics,
the total number of particles is $\sim 72$ on
the left part of the jump, and reaches $\sim 80$ on the most
right data point; the rotation frequency is increased step by step
from 0 to 0.4$\omega$, the converged state for a given $\Omega$
being taken as an initial guess in the iterative algorithm for
the successive value of $\Omega$.}
\label{fig:static}

\end{figure}

What is the physical origin of the gentle scenario ? The observed corrugation at
the surface suggests that it is driven by the instability of some surface mode
localized at the surface of the cloud, which is reminiscent of the Landau mechanism.
As a test of this idea, we have performed a numerical calculation of the
stationary BCS state in a rotating frame, by the above mentioned iterative scheme: 
as shown in figure \ref{fig:static} giving the angular momentum
of the stationary BCS solution as a function of the rotation frequency,
for $\epsilon=0.1$, the branch with no vortex is followed up to $\Omega=0.3\omega$;
for larger values of $\Omega$, the algorithm jumps to a configuration with vortices.
This suggests that the vortex free BCS state is indeed not a local
minimum of energy for $\Omega>0.3\omega$. What is then puzzling at this stage is that
a harmonic stirrer can excite only the quadrupolar modes, whereas the negative
energy surface mode initiating the Landau mechanism is expected to have
a higher angular momentum \cite{Dalfovo}.  A possible solution to this paradox 
was obtained by running the time dependent simulation with $\epsilon=0$, for $\Omega=0.5\omega$: 
in this case, the harmonic
trap, being isotropic, can not stir the gas and the stirring is due only to the
fixed periodic boundary conditions in the rotating frame; still, vortices were found to enter
the cloud. This shows that the quantization box in our $32\times 32$ 
simulations is small enough
to activate the Landau mechanism.

\noindent {\it Simulations on larger grids:}
To get rid of the previously mentioned finite quantization box effects,
we performed simulations on larger grids, $48\times 48$ and $64\times 64$.
For the $48\times 48$ grids, we investigated the rotation frequencies from
$0.4\omega$ to $0.8\omega$ in steps of $0.1\omega$. No vortex is found to enter the
cloud for $\Omega\leq 0.5\omega$. For $\Omega=0.6\omega$, vortices enter according to the
gentle scenario; the first vortices (in the form of a pair) enter however at a considerable later
time, $t\simeq 400\omega^{-1}$, than with the $32\times 32$ grid simulation.
For $\Omega\geq 0.7\omega$ the vortices enter according to the turbulent scenario.
The turbulent scenario on the $48\times 48$ grid
is similar to the one observed on the $32\times 32$ grid,
except for temporal shifts: e.g.\, on the $48\times 48$ grid,
the turbulent period starts $\simeq 240\omega^{-1}$ later for $\Omega=0.7\omega$
and $\simeq 65 \omega^{-1}$ later for $\Omega=0.8\omega$.

\begin{figure}[htb]
\begin{tabular}{cc}
\includegraphics[width=43mm,clip]{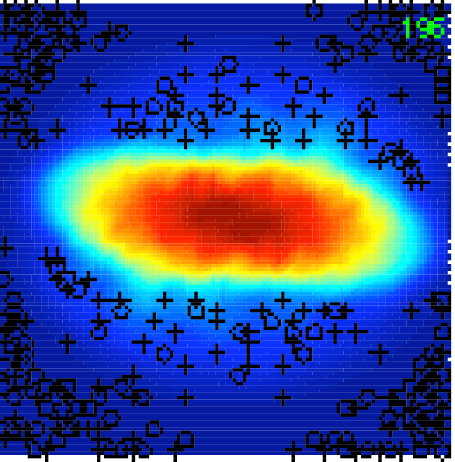} &
\includegraphics[width=43mm,clip]{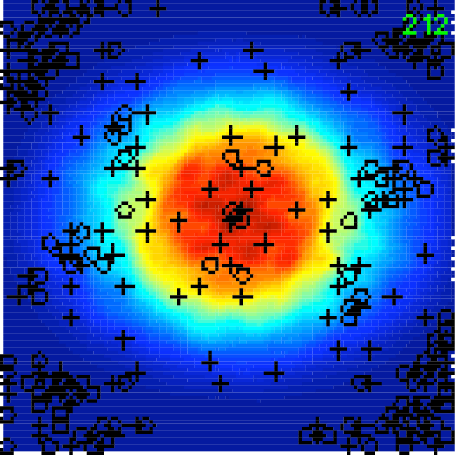} \\

\includegraphics[width=43mm,clip]{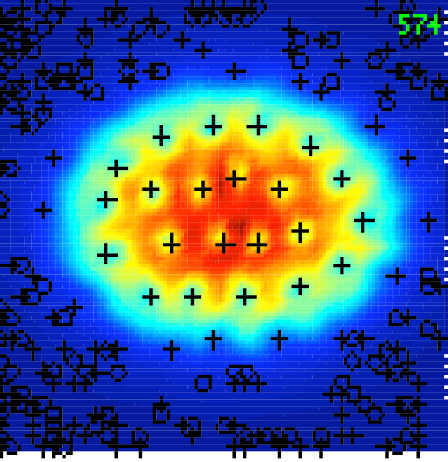} &
\includegraphics[width=43mm,clip]{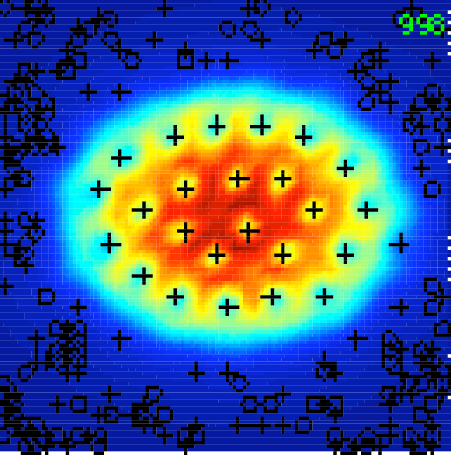} \\
\end{tabular}
\caption{For the numerical simulation of the 2D time dependent BCS equations
on a $64\times 64$ grid, density plots
of the density of the trapped gas at selected times (in units of $\omega^{-1}$), 
for a final rotation frequency $\Omega=0.8\omega$.
The trap anisotropy was $\epsilon=0.1$ and the 2D scattering length
$a_{2D}=\sqrt{\hbar/m\omega}$,
and $\mu=8\hbar\omega$ in the initial state.
The spatial width of the simulation is truncated in the figure to have approximately
the same width as in Fig.\ref{fig:movie_0.5}.
Crosses: positive charge vortices.
Circles: negative charge vortices.
From top to bottom and from left to right:
$t=196\omega^{-1}$: a turbulent, elongated cloud is formed;
$t=212\omega^{-1}$: the cloud is round again, and includes a disordered
mixture of vortices and anti-vortices; $t=574\omega^{-1}$: the vortices
crystallize in a quasi-stationary pattern; $t=998\omega^{-1}$: 
slow and small shifts of some vortex positions have taken place
with respect to the previous density plot.}
\label{fig:movie_0.8}
\end{figure}

For the $64\times 64$ grids, the CPU time for a single realization exceeds one month
on a bi-processor AMD Opteron workstation, so that we have considered only
two values of the rotation frequency.
For $\Omega=0.6\omega$, no entry of vortices is observed.
This confirms that the observation of the gentle scenario, at least up
to the maximal evolution time (1000$\omega^{-1}$) considered here, 
is an artifact of the quantization box.
For $\Omega=0.8\omega$, vortices enter according to the turbulent scenario.
The timing is now quantitatively the same at the $48\times 48$ simulation: in both
simulations, the vortices enter the cloud at $t\simeq 205\omega^{-1}$ and
crystallize in a quasi stationary pattern at times $\sim 500-550\omega^{-1}$.
Selected images of the movie of the $64\times 64$ simulation for 
$\Omega=0.8\omega$ are shown in figure \ref{fig:movie_0.8}.

\begin{figure}[htb]
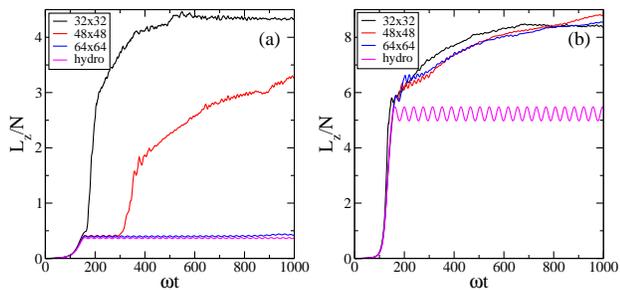

\begin{center}
\includegraphics[width=4cm,clip]{./comp_0.6.eps}
\includegraphics[width=4cm,clip]{./comp_0.8.eps}
\end{center}
\caption{Angular momentum per particle
in the gas, in units of $\hbar$, as a function of time,
for a final rotation frequency (a) $\Omega=0.6\omega$ 
and (b) $\Omega=0.8\omega$.
Curves (from top to bottom in (a)): numerical simulations of the 2D time dependent
BCS equations for the grids $32\times 32$, $48\times 48$ and
$64\times 64$, and for the time dependent superfluid hydrodynamic
theory of section \ref{sec:hydro}. 
}
\label{fig:comp}
\end{figure}

\begin{figure}[htb]
\begin{center}
\includegraphics[width=4cm,clip]{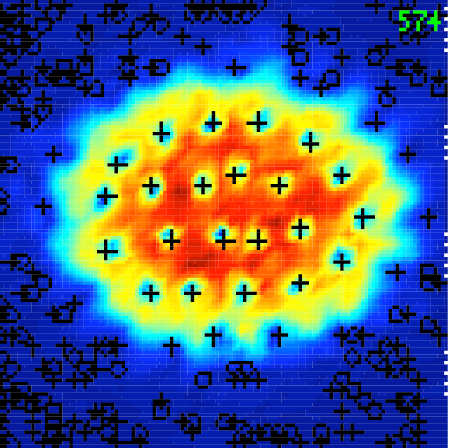}
\includegraphics[width=4cm,clip]{./extrait.eps}
\end{center}
\caption{At time $t=574\omega^{-1}$ of the $64\times 64$ numerical simulation
for $\Omega=0.8\omega$. Left panel: isocontours of the modulus of the gap
parameter, showing the presence of a vortex lattice; the $x$ and $y$
coordinates run from $-10a_{\rm ho}$ to $+10 a_{\rm ho}$ in the simulation
but this left panel figure is truncated to a position interval approximately
$-7 a_{\rm ho}$ to $+7 a_{\rm ho}$.
Right panel:
on the line $y=-0.627 a_{\rm ho}$, $x$ dependence of the density $\rho$
(solid line, in units of $a_{\rm ho}^{-2}$) and of the modulus of
the gap parameter (dashed line,
in units of $\hbar\omega$). The gap parameter was multiplied by 2/3 
for clarity.
A Fourier interpolation technique was used in the right panel
to map the 64$\times$64 simulation grid
onto a 128$\times$128 grid.}
\label{fig:cut}
\end{figure}

To allow for a quantitative comparison between the simulations
for the three grid
sizes, we have plotted in figure \ref{fig:comp} the total angular
momentum of the gas as a function of time,
for (a) $\Omega=0.6\omega$ and (b) $\Omega=0.8\omega$.
We have also given the (vortex free) hydrodynamic prediction;
remarkably,
this shows that the simulations give results close
to the hydrodynamic one as long as no vortex enters the cloud,
see in particular the $64\times 64$ results for $\Omega=0.6\omega$.
To briefly address the experimental observability of the vortex pattern,
we also show in figure \ref{fig:cut} a cut of the particle density (directly measurable
in an experiment) and of
the gap parameter (not directly accessible experimentally) for the 
$64\times 64$ simulation with $\Omega=0.8\omega$ at a time 
when the vortex lattice is crystallized,
this in parallel to an isocontour of the
magnitude of the gap parameter: vortices embedded in high density regions
result in dips in the density profile, with a contrast on the order here of 30\%.

\section{Conclusion}

We have investigated a relevant problem for the present experiments
on two-spin component interacting Fermi gases, the possibility to form a vortex
lattice by slow ramping of the rotation frequency 
of the harmonic trap containing the particles.
The observation of such a vortex lattice in steady state
would be a very convincing evidence of superfluidity \cite{vort_ketterle}.

For a 2D model based on the BCS theory, and for the 3D unitary quantum gas,
we predict analytically,
with the superfluid hydrodynamic equations, that the gas experiences
a dynamic instability when the final rotation frequency is above
some minimal value $\Omega_u$ that we have calculated. This dynamic
instability is very similar to the one discovered for a rotating Bose-Einstein
condensate of bosonic atoms, where it was shown to lead to the
vortex lattice formation.

To see if this dynamic instability leads to the formation of
vortices also in the case of the Fermi gases, we have solved numerically 
the full 2D time-dependent BCS equations, for a trap
anisotropy $\epsilon=0.1$.
For a final rotation frequency $\Omega$ above the predicted $\Omega_u$, we
see turbulence and the subsequent fast entry of vortices.
We conclude that the dynamic instability can indeed result
in a vortex lattice formation. The apparent irreversibility and energy dissipation
that this seems to imply may be surprising at first sight, since the equations of motion
that we integrated are purely conservative. The clue is probably the same as in
the bosonic counterpart of these simulations \cite{Sinatra}: the spatial noise
produced in the turbulent phase populates many eigenmodes (including collective modes)
of the system,
and the subsequent non-linear evolution leads to effective thermalization of the modes.

For $\Omega<\Omega_u$ but for $\Omega$ larger than what we estimated
to be the Landau rotation frequency (above which the vortex
free superfluid is no longer a local minimum of energy in the
rotating frame), we also see the formation of a vortex lattice in
the simulations on the small $32\times 32$ grids, 
but with a gentle mechanism not involving turbulence
and leading to the entrance of a pre-formed regular vortex pattern
from the surface of the cloud. 
But we also performed simulations on larger grids: on a
$64\times 64$ grid, the gentle mechanism disappears; 
it is therefore an artifact
of the periodic boundary conditions that rotate in the lab frame
and provide an artificial stirring effect.
In a real experiment, however,
we expect such a gentle mechanism to occur for a gas initially at finite temperature,
when the normal component of the gas is set into rotation by the stirrer.

\begin{acknowledgments}
We acknowledge useful discussions with C. Salomon, F. Chevy and A. Sinatra.
One of us (G.T.) acknowledges financial support from the European Union
(Marie Curie training site program QPAF).
Laboratoire Kastler Brossel is a Unit\'e de
Recherche de l'\'Ecole Normale Sup\'erieure et de l'Universit\'e Paris
6, associ\'ee au CNRS. 
\end{acknowledgments}

\appendix

\section{Simple derivation of the hydrodynamic equations from BCS theory}
\label{appen:hydro}

We show here that the time dependent hydrodynamic equations Eq.(\ref{eq:continuity})
and Eq.(\ref{eq:euler}) can be formally derived for a vortex free gas from the time dependent
BCS equations by using the lowest order semi-classical approximation and
an adiabatic approximation for the resulting time dependent equations.
As in the remaining part of the paper, we consider here the regime where the chemical
potential is positive and larger than the binding energy $E_0$.

The general validity condition of a semi-classical approximation is that the
coherence length of the gas should be much smaller than the typical 
length scales of variation of the applied potentials. Two coherence lengths 
appear for a zero temperature BCS Fermi gas: the inverse Fermi wave-vector,
$k_F^{-1}$, associated to the correlation function 
$\langle \psi^\dagger_\uparrow(\mathbf{r})\psi_\uparrow(\mathbf{r'})\rangle$,
and the pair size, $l_{\rm BCS} \sim \hbar^2 k_F/m |\Delta|$,
associated to the correlation function $\langle\psi_\uparrow(\mathbf{r})
\psi_\downarrow(\mathbf{r'})\rangle$. A first typical length
scale of variation of the matrix elements in Eq.(\ref{eq:uvt}) comes from
the position dependence of $|\Delta|$: in the absence of rotation,
we assume that this is the Thomas-Fermi
radius $R_{\rm TF}$ of the gas, defined as $\hbar^2 k_F ^2/2m = m\omega^2 R_{\rm TF}^2/2$.
This assumes that the scale of variation of the modulus of the gap is the same as the one of the density;
the adiabatic approximation to come will result in a 
$|\Delta|$ related to the density by Eq.(\ref{eq:delta_randeria}), which
justifies the assumption. Necessary validity conditions of a semi-classical approximation are then:
\begin{equation}
k_F^{-1}, l_{\rm BCS} \ll R_{\rm TF}.
\label{eq:cond_sc}
\end{equation}
In the BCS regime regime, $k_F^{-1} < l_{\rm BCS}$; for an isotropic
harmonic trap, one then finds that the condition Eq.(\ref{eq:cond_sc}) is
equivalent to
\begin{equation}
|\Delta| \gg \hbar \omega,
\label{eq:usual}
\end{equation}
where $\omega$ is the atomic oscillation frequency.
The presence of vortices introduces an extra length scale in the variation of $|\Delta|$,
on the order of $l_{\rm BCS}$, which invalidates the semi-classical approximation.

In the rotating case, however, this is not the whole story, as the phase of $\Delta$
can also become position dependent. As we shall see, the phase of $\Delta$ in this paper
may vary as $m \omega x y/\hbar$: when this quantity varies by $\sim 2\pi$, 
$\Delta$ changes completely; this introduces a length scale $\sim 2\pi\hbar/(m\omega R_{\rm TF})\sim
1/k_F$, making a semi-classical approximation hopeless. We eliminate this problem by performing
a gauge transform of the $u$'s and $v$'s: 
\begin{eqnarray}
\tilde{u}_s(\mathbf{r},t) &\equiv& u_s(\mathbf{r},t) e^{-i S(\mathbf{r},t)/\hbar} \\
\tilde{v}_s(\mathbf{r},t) &\equiv& v_s(\mathbf{r},t) e^{+i S(\mathbf{r},t)/\hbar}
\end{eqnarray}
where the phase is defined in Eq.(\ref{eq:delta}). The time dependent BCS equations are
modified as follows:
\begin{equation}
i\hbar \partial_t \, \left(\begin{tabular}{c} $\tilde{u}_s$ \\ $\tilde{v}_s$\end{tabular}\right) = 
\left(\begin{tabular}{cc} $\tilde{h}_0$ & $|\Delta|$  \\ $|\Delta|$ & $-\tilde{h}_0^*$ \end{tabular}\right) 
\left(\begin{tabular}{c} $\tilde{u}_s$ \\ $\tilde{v}_s$\end{tabular}\right)
\equiv \hat{L} \left(\begin{tabular}{c} $\tilde{u}_s$ \\ $\tilde{v}_s$\end{tabular}\right)
\label{eq:uvtm}
\end{equation}
where the gauge transformed Hamiltonian is
\begin{equation}
\tilde{h}_0= e^{-i S/\hbar} h_0 e^{+i S/\hbar} + \partial_t S.
\end{equation}

Let us review relevant observables in the gauge transformed representation. First the gap
equation is modified as
\begin{equation}
|\Delta| = - g_0 \sum_s \tilde{u}_s \tilde{v}_s^*.
\label{eq:deltam}
\end{equation}
Then the mean total density reads
\begin{equation}
\rho = 2\sum_s \tilde{v}_s \tilde{v}_s^*.
\label{eq:densm}
\end{equation}
Last, we introduce the total matter current $\mathbf{j}(\mathbf{r},t)$, that obeys by definition
\begin{equation}
\partial_t \rho + \mathrm{div} \, \mathbf{j} =0.
\label{eq:cont}
\end{equation}
In the rotating frame, in a many-body state invariant by exchange of the spin states
$\uparrow$ and $\downarrow$,
it is very generally 
given by 
\begin{equation}
\mathbf{j}=\frac{\hbar}{im}\left(\langle \psi_{\uparrow}^\dagger \mathbf{grad}\,\psi_{\uparrow}\rangle-
\mbox{c.c.}\right)
- \rho\, \mathbf{\Omega}\times \mathbf{r}.
\end{equation}
Within BCS theory, this gives
\begin{equation}
\mathbf{j}= \rho\, (\mathbf{v} -\mathbf{\Omega}\times\mathbf{r})
+\frac{i \hbar}{m} \sum _s \left[ \tilde{v}_s^* \, \mathbf{grad}\, \tilde{v}_s
-\tilde{v}_s \, \mathbf{grad}\, \tilde{v}_s^*\right],
\label{eq:j}
\end{equation}
where the velocity field $\mathbf{v}$ is defined as $\mathbf{grad}\,S/m$.
Note that the continuity equation Eq.(\ref{eq:cont}) remains true for
the BCS theory \cite{Blaizot}.

To calculate the two key quantities Eq.(\ref{eq:densm}) and Eq.(\ref{eq:j}), it is sufficient
to know the following one-body density operator for a fictitious particle of spin 1/2,
\begin{equation}
\sigma = \left(\begin{tabular}{cc}
$\sigma_{\uparrow\uparrow}$ & $\sigma_{\uparrow\downarrow}$ \\
$\sigma{\downarrow\uparrow}$& $\sigma_{\downarrow\downarrow}$
\end{tabular}\right) \equiv \sum_s \left(
\begin{tabular}{cc}
$ |\tilde{u}_s\rangle \langle \tilde{u}_s | $ & $ |\tilde{u}_s\rangle \langle \tilde{v}_s | $ \\
$ |\tilde{v}_s\rangle \langle \tilde{u}_s | $ & $ |\tilde{v}_s\rangle \langle \tilde{v}_s | $ 
\end{tabular}
\right).
\end{equation}

To prepare for the semi-classical approximation we introduce the Wigner representation of $ \sigma$
\cite{Wigner}:
\begin{equation}
W(\mathbf{r},\mathbf{p},t) = \mathrm{Wigner}\{\sigma\}\equiv
\int \frac{d^d\mathbf{x}}{(2\pi\hbar)^d} \langle \mathbf{r}-\mathbf{x}/2
|\sigma| \mathbf{r}+\mathbf{x}/2\rangle e^{i\mathbf{p}\cdot\mathbf{x}/\hbar}
\end{equation}
where $d$ is the dimension of space. 
The key observables have then the exact expressions:
\begin{eqnarray}
\label{eq:rho_wig}
\rho(\mathbf{r},t) &=& 2\int d^d\mathbf{p} \, W_{\downarrow\downarrow}(\mathbf{r},\mathbf{p},t) \\
\label{eq:delta_wig}
|\Delta|(\mathbf{r},t) &=& -g_0 \int d^d\mathbf{p}\, W_{\uparrow\downarrow}(\mathbf{r},\mathbf{p},t) \\
\label{eq:j_wig}
\mathbf{j}(\mathbf{r},t) &=& \rho \, (\mathbf{v} -\mathbf{\Omega}\times\mathbf{r}) \nonumber \\
&&- \frac{2}{m} \int d^d\mathbf{p}\, \mathbf{p}\, W_{\downarrow\downarrow}(\mathbf{r},\mathbf{p},t).
\end{eqnarray}
The semi-classical expansion then consists e.g.\ in
\begin{equation}
\mathrm{Wigner}\{V(\hat{\mathbf{r}})\sigma\} = [V(\mathbf{r})+\frac{i\hbar}{2}\partial_{\mathbf{r}}V
\cdot \partial_{\mathbf{p}}+\ldots ] W(\mathbf{r},\mathbf{p},t) .
\end{equation}
The successive terms we called zeroth order, first order, etc, in the semi-classical approximation.

We write the equations of motion Eq.(\ref{eq:uvtm}) up to zeroth order in the semi-classical
approximation:
\begin{equation}
i\hbar\partial_t W(\mathbf{r},\mathbf{p},t)|^{(0)}=  \left[L_0(\mathbf{r},\mathbf{p},t),W(\mathbf{r},\mathbf{p},t)\right]
\label{eq:zeroth}
\end{equation}
where the matrix $L_0$ is equal to
\begin{equation}
L_0(\mathbf{r},\mathbf{p},t)= \left(
\begin{tabular}{cc}
$\frac{p^2}{2m} - \mu_{\rm eff}(\mathbf{r},t) $ & $|\Delta|(\mathbf{r},t)$ \\
$|\Delta|(\mathbf{r},t)$ & $ - \frac{p^2}{2m}+\mu_{\rm eff}(\mathbf{r},t)$ 
\end{tabular}
 \right).
\label{eq:Lspin}
\end{equation}
We have introduced the position and time dependent function,
\begin{equation}
\mu_{\rm eff}(\mathbf{r},t) \equiv \mu -U(\mathbf{r},t) -\frac{1}{2} m v^2 +
m\mathbf{v}\cdot (\mathbf{\Omega}\times \mathbf{r}) - \partial_t S(\mathbf{r},t),
\label{eq:mu_eff}
\end{equation}
that may be called effective chemical potential for reasons that will become clear later.

At time $t=0$, the gas is at zero temperature. By introducing the spectral decomposition
of $\hat{L}(t=0)$ one can then check that
\begin{equation}
\sigma(t=0) =\theta[\hat{L}(t=0)]
\end{equation}
where $\theta(x)$ is the Heaviside function. Since $L_0(t=0)$ is the classical limit
of the operator $\hat{L}(t=0)$, the leading order semi-classical approximation
for the corresponding Wigner function is, in a standard way, given by
\begin{equation}
W(\mathbf{r},\mathbf{p},t=0) \simeq \frac{1}{(2\pi\hbar)^d}\,\theta[L_0(\mathbf{r},\mathbf{p},t=0)]
\end{equation}
that is each two by two matrix $W$ is proportional to a pure state $|\psi\rangle\langle \psi|$ with
\begin{equation}
|\psi(\mathbf{r},\mathbf{p},t=0)\rangle = \left(\begin{tabular}{c} $U_0(\mathbf{r},\mathbf{p})$ \\
$V_0(\mathbf{r},\mathbf{p})$
\end{tabular}
\right)
\end{equation}
where $(U_0,V_0)$ is the eigenvector of $L_0(\mathbf{r},\mathbf{p},t=0)$ of positive energy
and normalized to unity.
At time $t$, according to the zeroth order evolution Eq.(\ref{eq:zeroth}), each two by two
matrix $W$ remains a pure state, with components $U$ and $V$ solving
\begin{equation}
i\hbar\partial_t \left(\begin{tabular}{c} $U(\mathbf{r},\mathbf{p},t)$ \\
$V(\mathbf{r},\mathbf{p},t)$ \end{tabular} \right)
=L_0(\mathbf{r},\mathbf{p},t)
\left(\begin{tabular}{c} $U(\mathbf{r},\mathbf{p},t) $ \\
$V(\mathbf{r},\mathbf{p},t)$ \end{tabular} \right)
\end{equation}

We then introduce the so-called adiabatic approximation: the vector $(U,V)$, being
initially an eigenstate of $L_0(\mathbf{r},\mathbf{p},t=0)$, will be an instantaneous
eigenvector of $L_0(\mathbf{r},\mathbf{p},t)$ at all later times $t$.
This approximation holds under the adiabaticity condition \cite{adiab}, 
detailed below,
requiring that the energy difference between the two eigenvalues of $L_0(\mathbf{r},\mathbf{p},t)$
(divided by $\hbar$) be large enough.
As this energy difference can be as small as the gap parameter,
this will impose a minimal value to the gap, as we shall discuss later. In this adiabatic
approximation, one can take 
\begin{equation}
W(\mathbf{r},\mathbf{p},t)=\frac{1}{(2\pi\hbar)^d}\,\theta[L_0(\mathbf{r},\mathbf{p},t)]=
\frac{1}{(2\pi\hbar)^d}\,|+\rangle\langle +|
\label{eq:++}
\end{equation}
where $|+(\mathbf{r},\mathbf{p},t)\rangle$, of real components $(U_{\rm inst},V_{\rm inst})$, is the
instantaneous eigenvector with positive eigenvalue
of the matrix $L_0$ defined in Eq.(\ref{eq:Lspin}).
Its components are simply the amplitudes on the plane wave $\exp(i\mathrm{p}\cdot\mathrm{r}/\hbar)$
of the BCS mode functions of a spatially uniform BCS gas of chemical potential $\mu_{\rm eff}$
and of gap parameter $|\Delta(\mathbf{r},t)|$. Using Eq.(\ref{eq:rho_wig}) and Eq.(\ref{eq:delta_wig})
with the approximate Wigner distribution Eq.(\ref{eq:++}), 
one further finds that this fictitious spatially uniform
BCS gas is at equilibrium at zero temperature so that the expressions 
Eq.(\ref{eq:eos}) and Eq.(\ref{eq:delta_randeria}) may be used.
In particular, Eq.(\ref{eq:eos}) gives
\begin{equation}
\mu_{\rm eff}(\mathbf{r},t) = \mu_0[\rho(\mathbf{r},t)]
\label{eq:euler_obtenu}
\end{equation}
which leads, together with Eq.(\ref{eq:mu_eff}), to one of the time dependent hydrodynamic equations, the Euler-type one
Eq.(\ref{eq:euler}). 
Also, $U_{\rm inst}$ and $V_{\rm inst}$ are even functions of
$\mathbf{p}$, so that the integral in the right hand side of Eq.(\ref{eq:j_wig})
vanishes and Eq.(\ref{eq:cont}) reduces to the 
hydrodynamic continuity equation Eq.(\ref{eq:continuity}).
Under the adiabatic approximation, the superfluid hydrodynamic equations are
thus derived.

We now discuss the validity of the adiabatic approximation.
Without this approximation, the two by two matrix $W$ has non-zero off-diagonal
matrix elements $\langle +|W|-\rangle$ where $|-\rangle$ is the instantaneous eigenvector
of Eq.(\ref{eq:Lspin}) with a negative eigenvalue, that can be written $(V_{\rm inst},-U_{\rm inst})$.
Writing from Eq.(\ref{eq:zeroth}) the equation of motion for $\langle +|W|-\rangle$, one 
indeed finds a coupling
to the diagonal element $\langle +|W|+\rangle$ due to the non infinite ramping time of the rotation.
This coupling can be calculated using the off-diagonal Hellman-Feynman theorem
for real eigenvectors,
and corresponds to a Rabi frequency
\begin{equation}
\frac{1}{2} \nu_{\rm time} \equiv -\langle -|\partial_t |+\rangle= 
-\frac{1}{\epsilon_+-\epsilon_-}\langle -|\left(\partial_t L_0\right)|+\rangle
\end{equation}
where $\epsilon_\pm$ is the eigenenergy of $|\pm\rangle$ for the matrix $L_0$:
\begin{equation}
\epsilon_{\pm} = 
\pm \left[\left(p^2/(2m)-\mu_{\rm eff}\right)^2+|\Delta|^2\right]^{1/2}.
\end{equation}
But this is not the whole story, as we have neglected
the so-called motional couplings, that can also destroy adiabaticity.
These motional couplings are 
due to the fact that $|+\rangle$ and $|-\rangle$ depends on $\mathbf{r},\mathbf{p}$
and that terms involving $\partial_{\mathbf{p}} W$ and $\partial_{\mathbf{r}} W$
will appear in the equation for $W$ beyond the zeroth-order semi-classical approximation.
Such non-adiabatic effects are well known for the motion of a spin $1/2$ particle in a static
but spatially inhomogeneous magnetic field.
In our problem, the first order term of the semi-classical expansion is actually simple to write:
\begin{equation}
\partial_t W|^{(1)} = \frac{1}{2} \left[\partial_{\mathbf{r}} L\cdot\partial_{\mathbf{p}} W
-\partial_{\mathbf{p}} L\cdot\partial_{\mathbf{r}} W +\mbox{h.c.}
\right].
\end{equation}
The matrix $L$ corresponds to the classical limit of $\hat{L}(t)$:
\begin{equation}
L(\mathbf{r},\mathbf{p},t)=L_0(\mathbf{r},\mathbf{p},t)  
+\mathbf{p}\cdot\left(\mathbf{v}-\mathbf{\Omega}\times\mathbf{r}\right)\,
\mbox{I},
\end{equation}
where I is the $2\times 2$ identity matrix.
In the resulting equation of evolution of $\langle +|W|-\rangle$, taking $\langle +|W|+\rangle=
1/(2\pi\hbar)^d$
and $\langle -|W|-\rangle=0$,
a motional Rabi coupling to $\langle +|W|+\rangle$ now appears:
\begin{eqnarray}
\frac{1}{2}\nu_{\rm motion} \equiv&& -
\partial_{\mathbf{p}} \left[\mathbf{p}\cdot(\mathbf{v}-\mathbf{\Omega}\times\mathbf{r})\right]\cdot \langle-|\partial_{\mathbf{r}} |+\rangle
\nonumber \\
&& +\partial_{\mathbf{r}} \left[\mathbf{p}\cdot(\mathbf{v}-\mathbf{\Omega}\times\mathbf{r})\right] \cdot \langle-|\partial_{\mathbf{p}} |+\rangle.
\end{eqnarray}
Expressions similar to the one for $\langle -|\partial_t |+\rangle$ can be derived 
with the off-diagonal Hellman-Feynman theorem.

We now calculate the total Rabi frequency $\nu_{\rm tot}\equiv
\nu_{\rm time}+\nu_{\rm motion}$
at the local Fermi surface, that is for a value of the momentum such
that $p^2/2m=\mu_{\rm eff}(\mathbf{r},t)$. This is indeed at the Fermi surface that we expect
the adiabaticity condition to be most stringent, as the energy difference $\epsilon_+-
\epsilon_-$ takes there its minimal value, equal to twice the gap $|\Delta(\mathbf{r},t)|$.
Then $U_{\rm inst}=V_{\rm inst}=1/\sqrt{2}$ and the
expressions resulting from the Hellman-Feynman theorem are very simple:
\begin{equation}
\langle -|\partial_\lambda |+\rangle = 
-\frac{\partial_\lambda(\mu_{\rm eff}-p^2/2m)}{2|\Delta|},
\end{equation}
where $\lambda$ stands for $t$ or for an arbitrary component of the vectors $\mathbf{r}$ or $\mathbf{p}$.
We then get the condition for adiabaticity:
\begin{equation}
\frac{|\nu_{\rm tot}|}{2} =
\frac{1}{2|\Delta|} \left|
\frac{D\mu_{\rm eff}}{Dt}+ \left(\frac{\mathbf{p}\cdot\partial_{\mathbf{r}}}{m}\right)^2 S\right|
\ll 2|\Delta|/\hbar,
\end{equation}
where $D/Dt= \partial_t +\left(\mathbf{v}-\mathbf{\Omega}\times \mathbf{r}\right) 
\cdot\partial_{\mathbf r}$.

A fully explicit expression for $\nu_{\rm tot}$ can be obtained using the hydrodynamic
equations and taking the limit of a very long ramping time of the rotation, as is the case
in our simulations, so that the hydrodynamic variables are close to a steady state
and $S\simeq m \omega \beta(t) x y$.
Using Eq.(\ref{eq:euler_obtenu}) and the continuity equation Eq.(\ref{eq:continuity}),
one gets $D\mu_{\rm eff}/Dt = -\rho\mu_0'[\rho]\mathrm{div}\,\mathbf{v}\simeq 0$ so that
one is left with
\begin{equation}
\frac{1}{2}\nu_{\rm tot} = \frac{\beta(t) \omega p_x p_y}{m|\Delta|}.
\end{equation}
The constraint $|\nu_{\rm tot}/2|  \ll 2|\Delta|/\hbar$ then results in the condition
in 2D:
\begin{equation}
\hbar \omega \ll 4 E_0/|\beta(t)|,
\end{equation}
where $E_0$ is the dimer binding energy.
This condition is satisfied in our simulations as $\beta$ is at most $\sim 0.64$
(for $\Omega=0.8\omega$) and we took $a_{2D}=(\hbar/m\omega)^{1/2}$,
$\mu=8\hbar\omega$ resulting in
$E_0 \sim 1.3 \hbar\omega$ and $\Delta\sim 4.7\hbar \omega$. Note that 
it is in general more stringent than the usual
condition Eq.(\ref{eq:usual})
but for the particular parameters of our simulations, it turns
out to be roughly equivalent.

\section{A splitting technique conserving the mean number of particles}
\label{appen:split}

The standard splitting technique
approximates the evolution due to Eq.(\ref{eq:uvt}) during a small time step
$dt$ by first evolving the $(u_s,v_s)$ into $(u_s',v_s')$
with the kinetic energy and rotational energy during $dt$, and then
evolving the $(u_s',v_s')$ with the $\mathbf{r}$-dependent part of two by two matrix of Eq.(\ref{eq:uvt}) during $dt$,
for a {\sl fixed} value of $\Delta(\mathbf{r},t)=-g_0 \sum_s
u_s'(\mathbf{r}) v_s'^*(\mathbf{r})$.
This exactly preserves the unitary of the full
evolution, but the fact that a {\sl fixed } value of $\Delta$ is taken
during the second step of the evolution 
breaks the self-consistency between $\Delta$ and $u_s,v_s$ so that the total 
number of particles, 
$N=2\sum_s \langle v_s|v_s\rangle$, is conserved to first order in $dt$
but not to all orders in $dt$.
Numerically, for the time steps $dt$ leading to a reasonable
CPU time, one then observes strong deviations of
this total number from its initial value. 
Note that such a problem does not arise 
for the time dependent Gross-Pitaevskii equation for bosons, for which 
conservation of unitary and number of particles is one and a same thing.

This problem for the BCS equations
can be fixed by restoring the self-consistency for the evolution 
during $dt$ associated to the
$\mathbf{r}$-dependent part of the equation of evolution. That is 
one solves during $dt$:
\begin{equation}
i\hbar \partial_t \, \left(\begin{tabular}{c} $u_s$ \\ $v_s$\end{tabular}\right) =
\left(\begin{tabular}{cc} $U(\mathbf{r})-\mu$  & $\Delta(\mathbf{r},t)$  \\ $\Delta^*(\mathbf{r},t)$ & $\mu-U(\mathbf{r})$ \end{tabular}\right)
\left(\begin{tabular}{c} $u_s$ \\ $v_s$\end{tabular}\right)
\label{eq:loc}
\end{equation}
not for a fixed $\Delta$ but with the time dependent $\Delta$
given by the self-consistency condition Eq.(\ref{eq:gape}).
As a consequence, Eq.(\ref{eq:loc}) written for all modes $s$
is a set of non-linearly coupled time
dependent equations.
Fortunately, they are purely local in $\mathbf{r}$,
so that they can be solved analytically.
One finds that $\Delta(\mathbf{r},t)$ 
varies as $e^{-i\lambda(\mathbf{r}) t}$,
where 
\begin{equation}
\hbar\lambda(\mathbf{r}) = 2[U(\mathbf{r})-\mu] 
-g_0 \sum_s \left[|v_s(\mathbf{r},t)|^2-|u_s(\mathbf{r},t)|^2\right]
\end{equation}
can be checked to be time independent for the local evolution Eq.(\ref{eq:loc}). Then the system
Eq.(\ref{eq:loc}) is transformed into one with time independent coefficients (so readily integrable)
by performing a time dependent gauge transform, 
$u_s(\mathbf{r},t)=U_s(\mathbf{r},t) e^{-i\lambda(\mathbf{r})t/2}$
and $v_s(\mathbf{r},t)=V_s(\mathbf{r},t) e^{+i\lambda(\mathbf{r})t/2}$.


\begin{thebibliography}{99}
\bibitem{proviso}
We note that the chemical potential $\mu$ is a function of the rotation frequency $\Omega$,
since we are here in a system with a fixed total number of particles. We do not
need here the explicit expression for $\mu$.
\bibitem{to_be_complete}
In this figure, it is not apparent that the very narrow tongues actually touch the
$\epsilon=0$ axis; but this can be shown analytically,
and the contact points are $\Omega=\omega/\sqrt{n}$, which is the value
of the rotation frequency ensuring the resonance of the rotating anisotropy
with the surface
mode of frequency $\sqrt{n}\omega$ in the lab frame \cite{Subhasis}.
Note also that the degrees $n=6$ and $n=7$ (not shown in
the figure) have each an additional instability
zone, however with very low values of the Lyapunov exponent ($\sim 0.01\omega$) so
of little physical relevance. We have not explored $n\geq 8$.
\bibitem{how_to_get_that}
This value is obtained from the following result:
the boundaries of the crescent of degree $n=3$ can be parametrized
as $\epsilon=\pm z_0 (-z_0^2+2\Omega^2-1)/\Omega$
where $z_0$ is a root of the degree 5 polynomial
$P(z)=3 z^5+19 \Omega z^4 +(6-13 \Omega^2) z^3
+(26 \Omega -25 \Omega^3) z^2 + (3+3\Omega^4-10\Omega^2) z
+3\Omega^5-22\Omega^3 + 7\Omega$, $\Omega$ being expressed
in units of $\omega$.
\bibitem{mean_field} The mean field term is simply $g_0 \rho/2$. It tends to zero in
the limit of a grid spacing $l$ tending to zero, both in the 3D and 2D BCS theories.
In 2D however, this convergence is only logarithmic so it can never be achieved in
a practical numerical calculation. Keeping this
mean field term introduces a spurious dependence on $l$ in the BCS
equation of state. This, as we have checked numerically, leads to a spurious
dependence on $l$ of the density profile and the gap parameter for a stationary
gas in the trap in the BCS approximation: we have therefore removed the mean field
term. In an exact many-body solution of the lattice model, not relying on the Born approximation
as the Hartree-Fock mean field term does, we do not expect such a slow convergence
in the $l\rightarrow 0$ limit.
\end{thebibliography}
\end{document}